\documentclass[12pt,preprint]{aastex}

\slugcomment{To appear in the Astrophysical Journal: Version 2006 Nov 30}

\shorttitle{Detection of $^{13}C$ Isotopic Species of $HC_7N$}
\shortauthors{Langston and Turner}

\begin{document}

\newcommand{\PM}{{\ifmmode {\pm}\else $\pm$\fi}}
\newcommand{\vdag}{(v)^\dagger}
\newcommand{\myemail}{glangsto@nrao.edu}
\newcommand{\etal}{{\it et al.}\ }
\newcommand{\myitem}[1]{\item{\makebox[5cm][l]{\bf {#1}}}}

\newcommand{\HCTHREEN}{{$HC_3N$}}
\newcommand{\HCFIVEN}{{$HC_5N$}}
\newcommand{\HCSEVENN}{{$HC_7N$}}
\newcommand{\HCNINEN}{{$HC_9N$}}

\newcommand{\TMCN}{{TMC-HC$_9$N}}      
\newcommand{\TMCG}{{TMC-GB}}      

\title{Detection of $^{13}C$ Isotopomers of Molecule $HC_7N$}

\author{G. Langston\altaffilmark{1}}
\affil{National Radio Astronomy Observatory, 
P.O. Box 2, Green Bank, WV 24944}
\email{glangsto@nrao.edu}
\and
\author{B. Turner\altaffilmark{2} }
\affil{National Radio Astronomy Observatory, 
520 Edgemont Rd, 
Charlottesville, VA 22901}
\email{bturner@nrao.edu}


\begin{abstract}
The $^{13}C$ substitutions of molecule $HC_7N$ were observed in TMC-1 using
the $J = 12 - 11$, $J = 13 - 12$ rotational transitions in the frequency
range 12.4 to 13.6 GHz. 
We present the first detection the $^{13}C$ isotopic species of $HC_7N$ in the 
interstellar medium, based on the average of 
a number of weak rotational transitions.

This paper describes the calibration and data averaging process 
that is also used in a search for large cyanopolyyne 
molecules in TMC-1 using the 100m Robert C. Byrd Green Bank Telescope (GBT).   
The capabilities of the  GBT 11 to 15 GHz observing system are described 
along with a discussion of numerical methods for averaging observations of a number
of weak spectral lines to  detect new interstellar molecules.
\end{abstract}


\keywords{astrochemistry, ISM:molecules, \HCSEVENN, Taurus Molecular Cloud, TMC}


\section{Introduction}

The large cyanopolyyne molecules,  \HCFIVEN, \HCSEVENN, and
\HCNINEN, are seen in great abundance towards the 
Taurus Molecular Cloud  1 (TMC-1) over a very wide
frequency range (e.g. \citet{ola88}, \citet{bel97},  \citet{kai04}, \citet{kal04}).
The detection of these cyanopolyynes demonstrates that large
organic molecules are forming in interstellar (IS) medium.

A notable property of the cyanopolyynes is the
relatively high intensity rotational transition lines seen 
towards astronomical sources.
The strong line intensity is due to the simplicity of the
radio spectra of these  
linear carbon chains with large dipole moments.
The cyanopolyynes
are valuable guides for the study of all large molecule chemistry
in the IS medium.  
The chemical relationships and distribution in TMC-1
of molecules \HCTHREEN, \HCFIVEN, \HCSEVENN,
and \HCNINEN\ have been studied by many authors, including
\cite{pra97}, \cite{dic01} and \cite{tur05}.
The relative abundance of deuterated and $^{13}C$ isotopomers
of molecules in the IS medium is an important test for models of
IS chemistry \citep{tur01}.
 
In \S\ 2 we present the observations and the technique for calibration and 
averaging of the data. 
In \S\ 3 we combine multiple observations of different rotational transitions of a 
single molecule.
In \S\ 4 we compare our measurements with earlier 
observations.
In \S\ 5 the results are summarized.

\section{Observations}

We present selected observations of TMC-1 which are part of a search
for large cyanopolyynes ($HC_{11}N$ and $HC_{13}N$) in the interstellar medium. 
The large cyanopolyyne search is reported by \cite{lan07b}.
The data presented here are from daily system check 
observations to confirm proper operation of the antenna pointing, receiver
sensitivity and spectrometer configuration.
All the search observations were made in the period January 14, 2006 to September 10, 2006.
The observations were made using the NRAO
\footnote{The National Radio Astronomy Observatory (NRAO) is a 
facility of the National Science Foundation operated under 
cooperative agreement by Associated Universities, Inc.}
Robert C. Byrd Green Bank Telescope
(GBT).  
We present observations of \TMCG\
(R.A. $4^h41^m42.05^s, ~\delta~ 25\arcdeg41\arcmin27.5\arcsec$ J2000), 
and \TMCN\ 
(R.A. $4^h41^m44.7^s, ~\delta~ 25\arcdeg40\arcmin56\arcsec$ J2000).
Figure 1, from \citet{lan07a}, shows a map of the $HC_9N$ emission
with the locations of \TMCN\ and \TMCG\ marked.

For these observations, the GBT was configured using 
the Astronomers Integrated Desktop, ASTRID, for
use of the Ku-band (11.7 to 15.6 GHz) dual beam receiver.  
At 11.7 GHz the FWHM beam width is 65\arcsec\
and at 15 GHz the FWHM beam width is 50.8\arcsec.
The GBT Ku-band receiver has two feeds, separated by an angular offset of
330\arcsec\ in cross-elevation direction on the sky.    

Before each spectral line observing session, we first performed PEAK
and FOCUS observations on bright continuum radio source 3C123.
Unless the weather was very poor, the first PEAK and FOCUS observations
were almost always successful.   Typical pointing offsets were under $10\arcsec$
in either axis.
In the case of unusually large offsets ($d\theta > 20\arcsec$), found only during poor weather,
the PEAK observations were repeated.   

All spectral line observations were performed using the ASTRID NOD procedure.
The NOD procedure allows simultaneous "signal" and "reference" observations
using both beams of the Ku-band receiver.   
A NOD observation consists of two 4-minute scans, one with the first beam of the
dual beam receiver pointing towards the target location, followed by a
second scan with the second beam pointed towards the target location.
The Ku-band receiver is oriented so that the two beams observe the same target
elevation.
During the spectral line observations, the GBT spectrometer is configured
to simultaneously collect spectra at the same frequencies in both beams.
For these observations, the spectrometer continuously controlled injection
of a 2-3 K noise diode signal into both beams via RF inputs of both
polarizations.   
The noise diode was turned on and off at a 1 Hz rate.

For the $HC_{13}N$ search observations, the GBT IF path and spectrometer were
configured to simultaneously collect four 50 MHz wide spectral bands centered
on four different frequencies.   The integration time was 30 seconds and the
scan duration was 4 minutes, yielding  8 spectra for each of the two polarizations
of the dual beams.   
The noise diode ON and noise diode OFF spectra were separately recorded.
During a four minute scan, a total of 
$8 (integrations) \times 2 (beams) \times 2 (polarizations) \times 2 (noise~diode~states) = 64 $
spectra were recorded.
The GBT spectrometer has 3 level and 9 level sampling modes.   
The 3 level sampling mode was used, yielding a channel spacing
$\Delta \nu =$ 3.05176 kHz.    
At 13 GHz this spectral resolution 
corresponds to 0.07 km/sec channels.

To confirm proper operations
of the GBT spectrometer, at the beginning of most sessions
we performed test NOD observations
on TMC-GB, simultaneously observed 
four strong cyanopolyyne rotational transition
lines, \HCFIVEN\ $J = 5 - 4$, 
\HCSEVENN\ $J = 12 - 11$, 
\HCSEVENN\ $J = 11 - 10$ and
 \HCNINEN\ $J = 23 - 22$, 
at frequencies listed in table 1.
In the short (4 minute) pair of NOD observations, all four lines
are visible at high signal to noise ratio. 

In the course of the large cyanopolyyne search, a map of TMC-1 was made in
the $HC_9N$ lines with 1 arc-minute angular resolution.
The intensity versus position maps are
described in \citet{lan07a}.    
The $HC_9N$ peak emission is
16 \% stronger than at the TMC-GB location.
This \TMCN\ position was observed
to confirm the detection isotopomers  of $HC_7N$ at TMC-GB.
This position was observed for a total of 4 hours on September 6 and 2 hours on 
September 7, 2006.   
The GBT spectrometer was configured for high resolution, 12.5 MHz 3 level samplers,
four spectral band, dual polarization observation.

For the 3 level 12.5 MHz bandwidth spectrometer configuration, the channel
spacing is 4 times smaller than the 50 MHz configuration,
$\Delta \nu =$ 0.76294 kHz.
At 13 GHz this spectral resolution 
corresponds to 0.018 km/sec channels.

For the 12.5 MHz spectrometer mode, two different IF input configurations were 
required for observation of all seven $^{13}C$ isotopomers of $HC_7N$
for the $J = 12 - 11$ rotation transitions, plus the $HC_7N$, $DC_7N$ and 
$HC_7$$^{15}N$ transitions.         

\subsection{$T_A$ Calibration}

The raw data were calibrated to an antenna temperature scale by the 
$T_{sys, reference} \times (C_{signal}(\nu) - 
C_{reference}(\nu)) / C_{reference}(\nu)$
technique, where $C_{signal}(\nu)$ is the measured intensity (raw counts) as
a function of frequency, $\nu$, when observing the source location, and 
$C_{reference}(\nu)$ is the measured intensity (raw counts) when observing one of
the two  reference locations.
The system temperature was
computed for each spectral channel of each reference location.   
The system temperature for each channel was defined as 
average of the noise diode ON and noise diode OFF  spectra divided by the 
difference of noise diode ON  and noise diode OFF spectra, then multiplied by
the effective temperature of the calibration noise diode value measured in the laboratory.
For the purposes of intensity calibration,
the reference system temperature,  $T_{sys, reference}$ is defined as the median 
of  the calculated system temperatures for the central 40 MHz of the 50 MHz 
spectral band.   This procedure was repeated for each of the four spectral bands. 
For the 12.5 MHz bandwidth observations the central 10 MHz of each spectral
bands was used for $T_{sys}$ calibration.

The two Ku-band beams are separated on the sky by an
amount smaller than the maximum size of TMC-1.   
To compensate for line contamination of the source spectrum due to possible 
emission in the reference spectra,
the calibration was performed using
a median filtered reference spectrum.
The median reference spectrum is a smoothed version of the
observed reference spectrum.  The median spectrum is the median
of all channels of the reference spectrum within a 0.5 MHz bandwidth centered on
the frequency of the input channel.
The filter width is wider than the 0.021 MHz FWHM TMC-1 line width.    

Figure 2 shows  
uncalibrated intensity versus frequency spectra, the signal beam spectrum (top), 
the reference beam spectrum (bottom).   
The frequency axis is set assuming a LSR source velocity of 5.815\PM 0.010 km/sec.
The 0.5 MHz wide median filtered (smooth) signal and reference spectra
are overlaid on the observed spectra.
The high effectiveness of median filtering in removing spectral lines is shown
for the median of the signal spectrum in figure 2.

The median width must be 
narrow enough to preserve significant variations in the IF gain as
a function of frequency,
but must be wide enough to reduce the noise in the
reference spectrum and to reject any weak narrow lines
at the reference position.
Calibration using the median spectrum produces a
higher signal to noise ratio spectrum than calibration without smoothing 
the reference spectrum.

\subsection{$T_B$ Calibration}
The $T_A$ scaled data were further calibrated to brightness temperature 
(corrected for atmospheric opacity) in order to compare these results with
previous observations and computing the physical properties of TMC-1.
The $T_B$ calibration was performed assuming TMC-1 is more extended than
the GBT primary beam in the frequency range 11.6 to 14.5 GHz.
This assumption is confirmed by our $HC_9N$ mapping observations.
  
The GBT beam efficiency was computed assuming
\begin{equation}
\eta_A(\nu) = 0.71 \exp[ -(\frac{\nu}{61.3 ~ GHz})^2]
\end{equation}
and
\begin{equation}
\eta_B(\nu) = 1.37~ \eta_A(\nu)
\end{equation}
where $\nu$ is the frequency in GHz.   Equation 1 is the
Ruze equation \citep{ruz66} assuming the GBT surface accuracy is 390 microns RMS
\footnote{http://wwwlocal.gb.nrao.edu/gbtprops/man/GBTpg/} and
$\eta_A(\nu=0)$ = 0.71.

For each observing epoch, the average 
atmospheric opacity, $\tau$, at 13.0 GHz was adopted for all all observed
frequency bands.   The adopted $\tau$ values ranged from 0.01 to 0.04,
with a median value of $\sim0.019$.
The $\tau$ values were computed based on a model for
the local weather (R. Madallena 2006
\footnote{http://www.gb.nrao.edu/$\sim$rmaddale/Weather/}),
using measured weather data at nearby locations.

Figure 3 shows the $T_B$ calibrated average spectrum
of the $HC_5N$ $J = 5 - 4$ line.   
The upper (black) spectrum shows the average of right and
left circular polarization using the GBT beam efficiency model of equation
2 and corrected for atmospheric opacity.
The two lower spectra are right (red) and left (blue) circular
polarizations, each $T_A$ calibrated.  
The GBT efficiency factor, $\eta_B$, is 0.926 at 13535 MHz.
For the median opacity, $\tau=0.019$, the
opacity correction factor is small, 1.02.

A spectral baseline is computed for each polarization separately.
The spectral baseline is the median of the observed spectrum
computed over a 0.5 MHz width centered on each channel of the
spectrum.  
A median baseline is commonly used
for GBT spectral line observations, which cover a relatively wide
bandwidth \citep{hol04}.
The near identical intensity of the left and right
circular polarization shows the good agreement of the 
two independent calibrations of the intensity of the two polarizations.
The $HC_5N$ $J= 5 - 4$ line was chosen as the
calibration example
to illustrate the importance of using a wide median filter window
for baseline subtraction, 
to preserve the line intensity in the case of a number of
closely spaced transistions.
The $HC_5N$ hyperfine
transitions are sufficiently close that a narrower median filter width in
the baseline calculation would reduce line intensity. 

The other 3 rotational transitions observed each epoch are
similarly calibrated.
Figure 4 shows  $T_B$ intensity versus frequency plot for
the four 50 MHz bands observed with the
GBT spectrometer, 
when configured for the four frequency band and dual polarization test mode.
Figure 5 shows the $T_B$ calibrated spectra for all four lines
in an intensity versus velocity plot.  

\subsection{Frequency Smearing due to Doppler tracking}
The GBT local oscillator (LO) and intermediate frequency (IF) system 
is designed with one doppler tracking system.   
This has important
consequences for observations using the GBT spectrometer in
the four frequency band mode.   The GBT system is designed
to accurately track the reference frame for the center frequency of the first of the 
four bands.   
All other up and down frequency conversions are
done in topocentric frame.    The consequence of this design
is that the observations for the first frequency band may
be averaged for long durations, without frequency smearing 
of the molecular lines.   For the remaining three frequency bands,
care must be taken to properly align the spectra before averaging
over several epochs.   We have written a number of 
{\it Interactive Display Language}
(IDL) \footnote{Research Systems, Inc. {\tt http://www.rsinc.com/idl}}
scripts
to properly account for the shift in center frequency of the spectral
bands over time.  
The data were reduced using  {\tt GBTIDL} 
\footnote{
Developed by the National Radio Astronomy Observatory, documentation
at http://gbtidl.sourceforge.net/}
and {\tt rt\_idl} 
\footnote{Documentation on {\tt rt\_idl} is available on the web at 
{\tt http://www.gb.nrao.edu/$\sim$glangsto/rt\_idl}}, 
data reduction packages.
During the observations, the data were examined using {\tt rt\_idl}
a set of display scripts created by G.L. for real-time display.  
A goal of the present work is to demonstrate
the effectiveness of these procedures.

Our process of averaging spectra with different channel bandwidths and center
frequencies is similar to the standard process for convolving irregularly sampled
continuum intensity values onto an image grid
\citep{dic02}.
The process of combining the spectra, $S_{e}(\nu)$, from each epoch of
observation has three major components.
Before averaging, each of the $S_{e}(\nu)$ spectra are $T_B$ calibrated.

\begin{description}
\myitem{Convolving function}
The first step in computing an average
spectrum from a set of spectra from each epoch, $S_{e}(\nu)$, with different 
center frequencies and
channel separation is definition of the output spectral sum, $Sum(\nu)$, and the convolving
function $W(\nu)$.   
For spectral gridding, convolution function, $W(\nu -\nu_0)$, 
is defined in equation 3 as a Gaussian function with FWHM width equal to 
half the channel width
of the first input spectrum, $\Delta \nu_0$ (for this application, all spectra
had nearly identical channel spacings).   The offset between the input
frequency channel of a single observation and the frequency of a channel 
of the average spectra is 
$\Delta \nu = \nu - \nu_0$. 
\begin{equation}
W(\Delta\nu) = \exp{[-4~ln(2)(\Delta\nu/\Delta \nu_0)^2]}
\end{equation}

This convolving function definition has three desired properties.   The first is
at zero frequency  offset the convolving function is unity, i.e. $W(\Delta\nu=0) = 1$.
The second property is if the offset is half the channel width, then
the two adjacent channels are averaged, i.e. $W(\Delta\nu = \pm \Delta\nu_0/2) = 0.5$.
The third property is if the input channels have a large frequency offset, then
these channels make little contribution to the average, 
i.e. $W(\Delta\nu = \pm 3 \Delta\nu_0) \sim 1.5 \times 10^{-11}$.
For our application, the weighting function kernel width is \PM5 channels.   

\myitem{Convolution}  The spectra from each epoch is gridded into the output spectrum
in the same manner.  For each channel in the output spectrum (frequency $\nu$),
the closest channels in the input spectrum were found (frequency $\nu_i$).    
Next over the convolution function kernel size, the frequency offset between
the input spectrum and the output channel was computed.  The
convolution function was computed and
multiplied by a constant spectral weight for each epoch of observation.   
The spectrum weight is $W_e =  t_{int}/T_{sys}^2$, where
$t_{int}$ is the integration time and $T_{sys}$ is the system temperature.
This weighting 
yields the optimum signal to noise ratio in the output spectrum. 
The  input spectrum was multiplied by the weighted convolution kernel and
summed into the output spectrum.
The weighted convolution kernel is summed into the convolution
weight spectrum.  

\begin{equation}
Sum (\nu) = \sum_{epochs} \sum^{+5}_{i = -5} W_e W(\nu - \nu_i) S_e(\nu_i)
\end{equation}
\begin{equation}
Weight (\nu) = \sum_{epochs} \sum^{+5}_{i = -5} W_e  W(\nu - \nu_i)
\end{equation}
\myitem{Normalization}
The output spectrum is produced from the ratio of the convolved 
weighted input values and the summed convolution function.   
\begin{equation}
Spectrum (\nu) = Sum(\nu) / Weight(\nu)
\end{equation}

\end{description}

\section{Combination of Multiple Transitions}
Both weak and strong lines may be averaged using this technique.
Figure 6 shows the weighted average of $HC_7N$ $J = 12 - 11$ and 
$J  = 11 - 10$ transitions, along with the individual transitions offset
by 0.5 K.   Table 2 shows the measured intensities for
the weighted average of the lines.

For molecules with simple structure, such as the cyanopolyynes, the
line strength of closely spaced J rotational transitions changes smoothly
and predictably.     
The line intensities are weak for larger interstellar molecules, and
several days of observations are required to detect these transitions.
For molecules with several closely spaced, similar strength, rotation transitions,
the total observing time to required to obtain a desired signal
to noise ratio is reduced linearly with the number of transitions observed.
For weak molecular line detections, we have demonstrated (as expected)
that the signal to noise ratio of the detection is doubled by averaging
observations of four transitions.
Larger linear molecules have many closely spaced, similar
strength transitions, so no weighting for expected line strength is required.

We use the frequencies measured in the laboratory by \citet{mcc00} to average
observations of different $^{13}C$ isotopic species of $HC_7N$.
Figure 7 shows intensity versus velocity plot for six $HC_7N$ isotopomers observed
daily.   Figure 8 shows the intensity versus velocity plot for the system temperature and
integration time weighted average. fit with a Gaussian model for the line intensity.
The measured intensity of the $HC_7N$ and isotopomers is given in Table 2.

To confirm the detection of the isotopomers, we performed a second observation
towards \TMCN.
Figure 9 shows the $HC_7N$ $J = 12 - 11$ rotational transition for the
observations of \TMCN.
Figure 10 shows the weighted average of seven $^{13}C$ isotopomers of
$HC_7N$ along with the individual lines, each offset by 0.1 K.
Figure 11 shows the weighted average and a Gaussian fit to the average line.
No significant emission was found at the frequency of $DC_7N$ $J = 12 - 11$, 
with a $3 \sigma$ upper limit of  0.019 K.
No significant emission was found at the frequency of the $^{15}N$ 
isotopomer  of $HC_7N$ $J = 12 - 11$, with a $3 \sigma$ upper limit of 0.019 K.

\section{Comparison with Earlier Observations}
\citet{bel97} observed TMC-1 with the NRAO 140ft (43m) telescope and 
measured the brightness of the $HC_7N$ $J = 12 - 11$ transition.  
For this transition, they found $T_B = 0.475 \PM 0.013$ K.
This value is $\sim 50$ \% of the intensity measured with the GBT.
At 13.5 GHz, the 140ft telescope FWHM beam width is  $\sim 130$ \arcsec.
TMC-1 covers a smaller fraction of the 140ft telescope beam area.
As figure 1 shows, TMC-1 is a roughly linear structure at the 140ft angular
resolution, so we expect to measure a greater intensity with the GBT,
roughly in relationship to the ratio of the GBT and 140ft telescope diameters.

\citet{tur01} observed the $DCN/HCN$ and $H^{13}CN/HCN$ abundance 
ratio for three interstellar clouds.
For TMC-1, they use two radiative transfer models to estimate
abundances.  
For their model "C" they found $DCN/HCN \sim 1/56$ and 
$H^{13}CN/HCN \sim 1/100$.
These values are consistent with the average value we find for 
the isotopomers of $HC_7N$.   Based on the observation of TMC-GB
we find the abundance ratio $^{13}C/C$ is 1/87 with $1 \sigma$ range
1/122 to 1/68.
Based on the observation of \TMCN\, the  
abundance ratio $^{13}C/C$ is 1/132 with $1 \sigma$ range
1/176 to 1/106.
Averaging the measurements from the two positions yields 
abundance $^{13}C/C$ is 1/108 with $1 \sigma$ range
1/136 to 1/90.

\citet{tak98} observed the $^{13}C$ isotopomers of $HC_3N$,
and $HC_5N$ in TMC-1.   
They find significant variation in the abundance ratios
of the different isotopomers of $HC_3N$.    
For the TMC-1 they find  $H^{13}CCCN/HC_3N = 79\PM11$,
$HC^{13}CN/HC_3N   = 75\PM10$,
$HCC^{13}CN/HC_3N = 55\PM7$.
Our average isotopic abundance is consistent with their
results, but our observations do not have sufficient sensitivity to
measure small differences in abundances of the different isotopic
species.

For deuterium, our $3 \sigma$  upper limit on the abundance ratio for 
$DC_7N/HC_7N$ is 1/59.
\cite{tur01}  has modeled the $^{13}C$, deuterium and $^{15}N$ 
isotopomers of $HCN$ and other IS molecules.
For molecule $HCN$ they find a $DCN/HCN$ ratio range of 1/100 to 1/55,
depending on the radiative transfer model assumed.   
This range is consistent with our limit.

\cite{tur01} adopt abundance ratio of $HC^{15}N/HCN$ of 1/500.    
We find an upper limit for the abundance ratio of 
$HC_7$$^{15}N/HC_7N$ $ < 1/52$  ($3 \sigma$).

It should be noted that for larger molecules, containing many
atoms, the isotopic species are a significant
fraction of the total molecular density.  
In the case of location TMC-GB, 
we find that the $^{13}C$ isotopomers of $HC_7N$
are $7\times 0.0099/0.864 \sim 8\%$ of the total molecular abundance.

\section{Summary}

We detect the $^{13}C$ isotopomers of $HC_7N$ and place upper limits
on $DC_7N$ and $HC_7$$^{15}N$ abundances.

We have demonstrated the benefits of averaging a number of weak
rotational transitions of a molecule based on
accurate laboratory measurement of spectral lines frequencies.

The GBT receiver, IF system and GBT spectrometer are great tools for 
observation of weak narrow molecular lines.   The flexibility of the
of configuring the four spectral bands is valuable
for maximizing observing efficiency.   
However further expansion of the GBT IF and spectrometer
are needed to fully observe all isotopic species of $HC_7N$
in the GBT Ku-Band receiver frequency range.
Currently we can simultaneously observe only 6 of the 42 transitions
in this frequency range.  
A factor of 7 improvement in observing efficiency is possible for this project by 
expanding the spectrometer's spectral coverage.   
Enhancing the spectrometers capabilities should be a priority for
detecting the many weak rotational transitions
expected for biologically significant molecules.
  
\acknowledgments
We thank P. Thaddeus, S. Breunken, C. Gottlieb, and M. McCarthy
of the Harvard Center for Astrophysics for pointing out the possibility
of detecting the $HC_7N$ isotopomers.
We thank Elizabeth Russell for help in organizing the data reduction
effort.  
Thanks to R. Maddalena and T. Minter for helpful discussions on GBT
calibration parameters.
Thanks  to F. Lovas for computing the \HCFIVEN\ hyperfine transition
frequencies.
We thank the GBT operations and engineering staff for
support during these observations.
We acknowledge the very helpful suggestions of M. Hollis.



{\it Facilities:} \facility{NRAO}



\appendix

\begin{figure}
\epsscale{.80}
\plotone{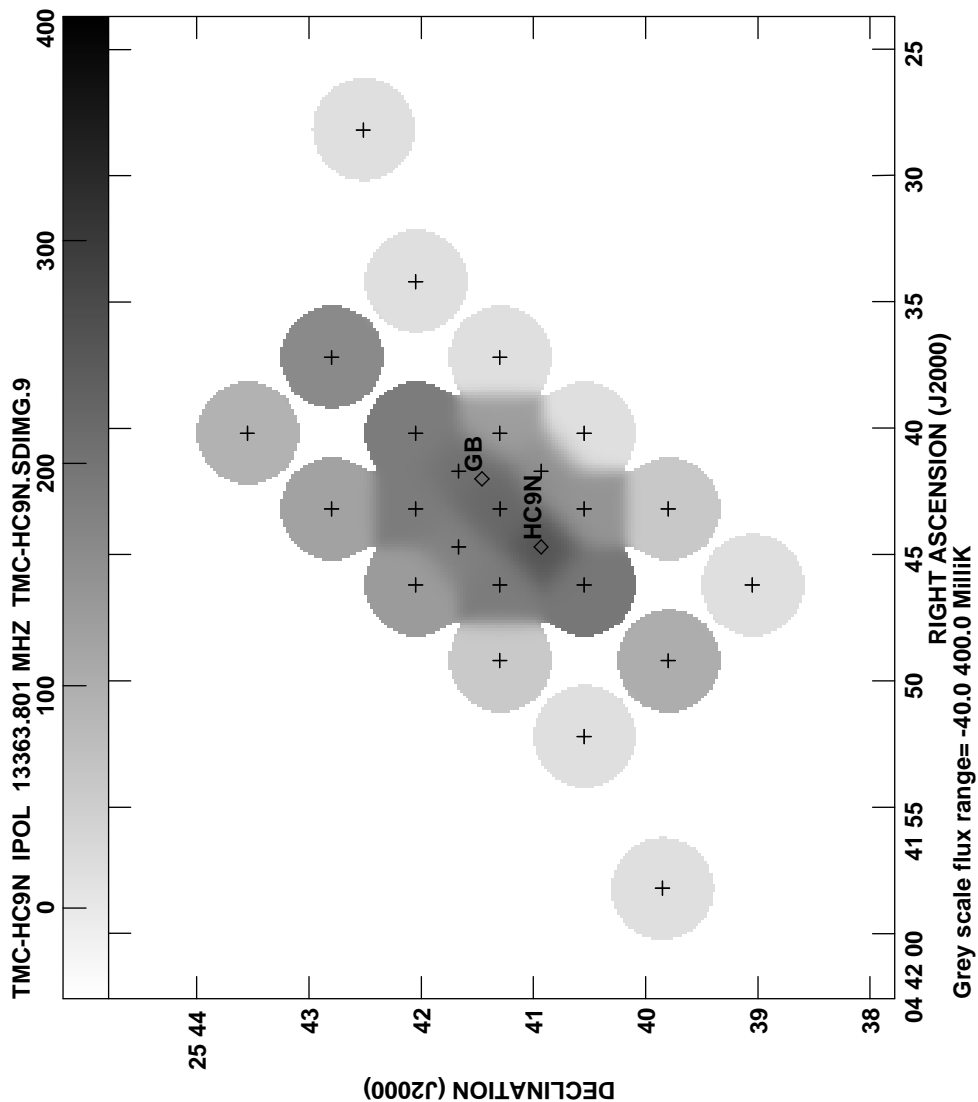}
\caption{$HC_9N$ molecular line intensity version position map of the TMC,   
from \citet{lan07a}.
The intensity scale is based on GBT observations of $T_B$ calibrated 
average intensity of the 
$HC_9N$ $J=22-21$, $J=23-22$, $J=24-23$ and $J=25-24$ rotational
transitions.  The location of the telescope pointings are shown with crosses.
The image convolving function width is the 55\arcsec.
The locations of \TMCG\ and \TMCN\ are marked with
diamonds.
}
\end{figure}

\begin{figure}
\epsscale{.80}
\plotone{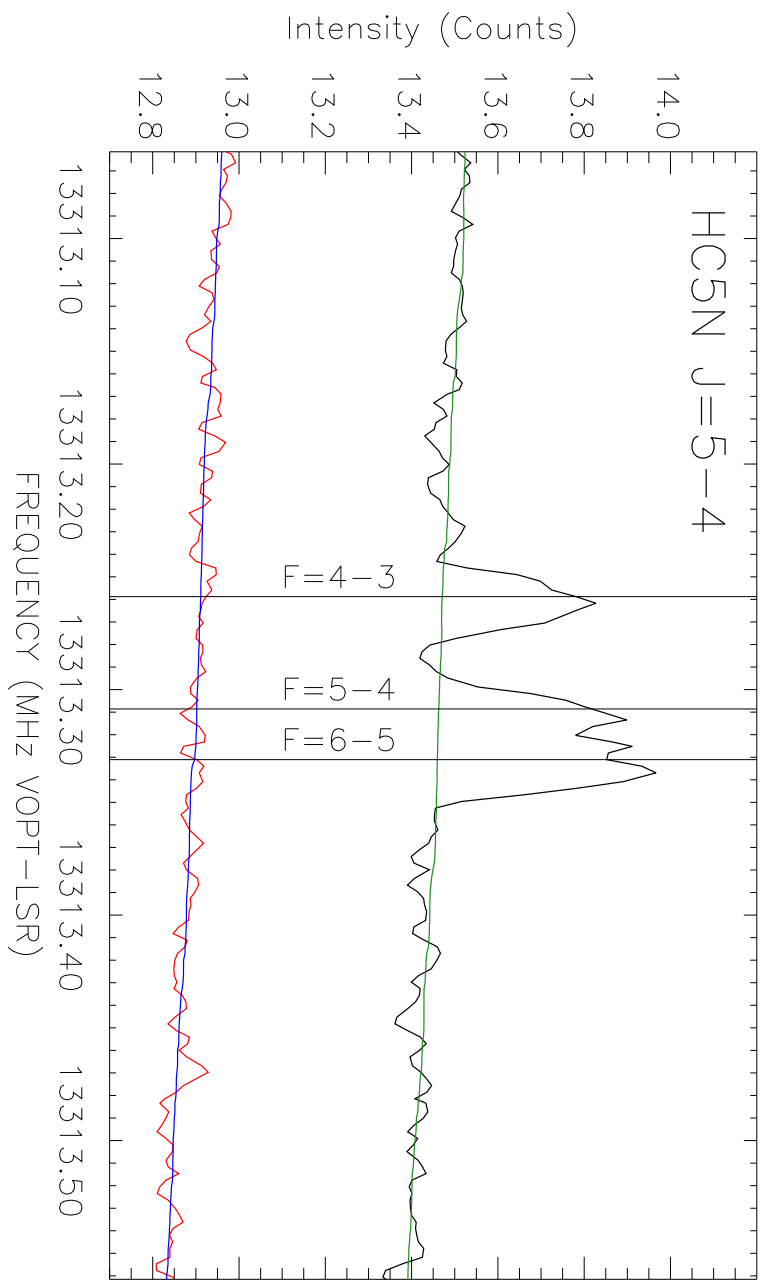}
\caption{Intensity versus frequency plots centered
on the three hyperfine transitions of the $HC_5N$  $J = 5 - 4$ line.
The upper spectrum (black) is the average of the signal beam observations, 
when pointed towards \TMCG.    
The lower spectrum (red) is the average of the reference 
beam observations, which are offset 330 \arcsec\ 
in azimuth from the signal beam location.
Overlaid on the reference beam spectrum is the median filtered reference
spectrum (blue), used for calibration of the signal spectrum.  The median
filter width is 0.5 MHz.
For comparison, the median filtered signal spectrum (green) is also shown.
For sufficiently wide filter window sizes, the molecular lines are completely
removed.
}
\end{figure}

\begin{figure}
\epsscale{.80}
\plotone{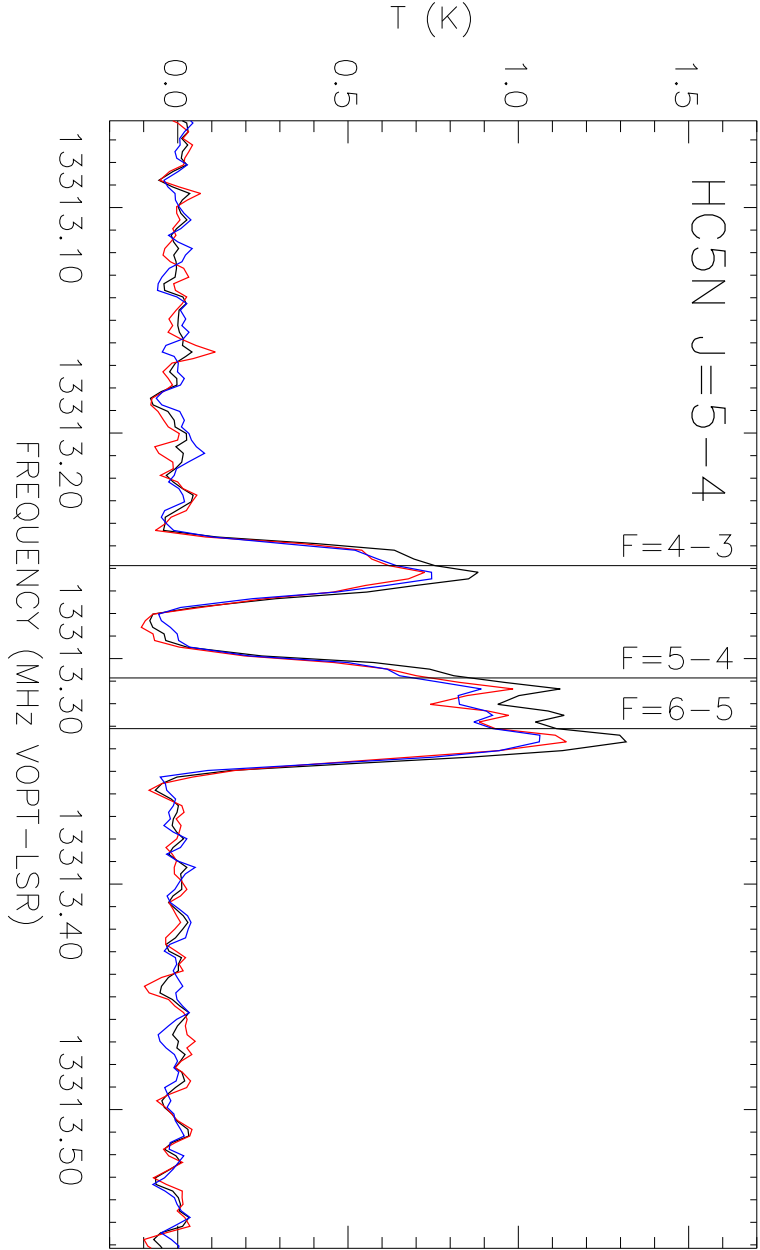}
\caption{Three intensity versus frequency spectra, showing the fully $T_B$ calibrated
average spectrum (black) from a NOD observation of TMC-GB.  
The red spectrum is right circular polarization (RCP) and the blue spectrum is the 
left circular polarization (LCP).  
The RCP and LCP spectra are on the $T_A$ scale, which does not include
corrections for beam efficiency and atmospheric attenuation.
A baseline has been subtracted from all three spectra, computed
from a 0.5 MHz wide median filter
of the original spectra.
These spectra are from 3.8 minutes of observations on 2006 May 12. 
Three hyperfine transition frequencies are marked.
}
\end{figure}

\begin{figure}
\epsscale{.80}
\plotone{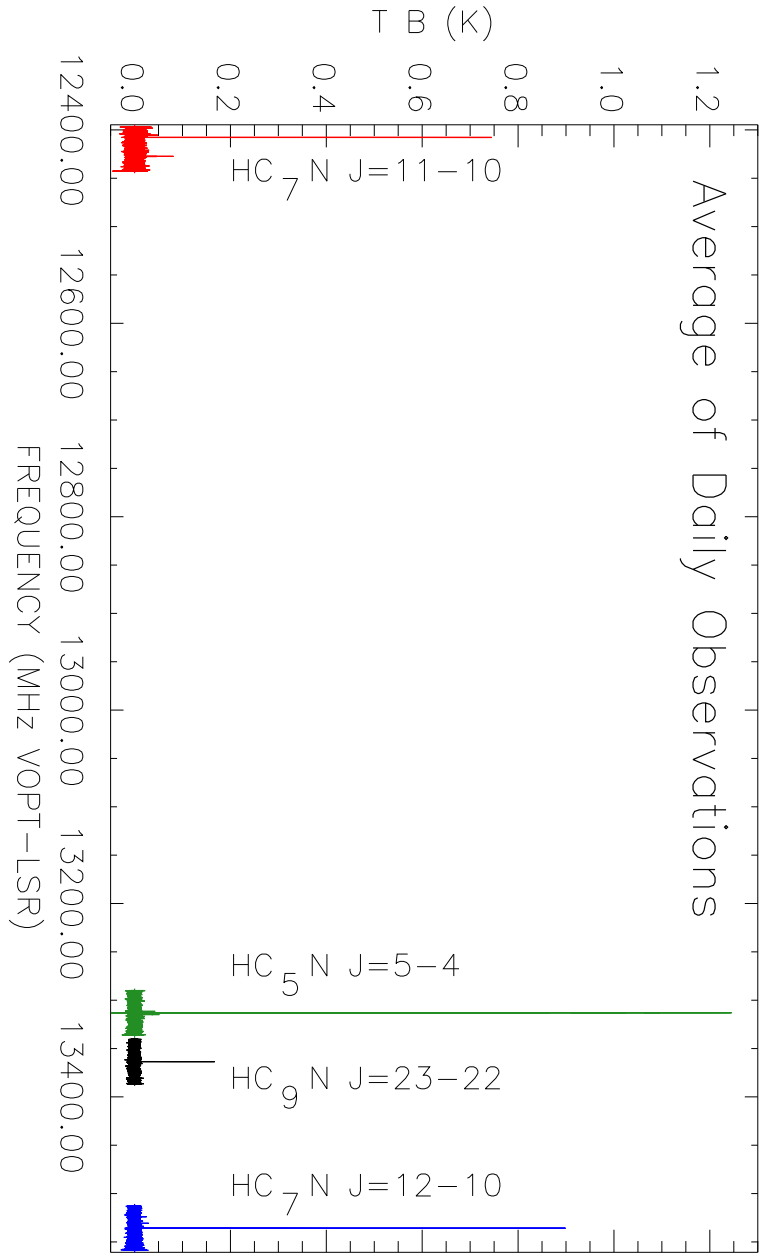}
\caption{Intensity versus frequency plot showing daily system check observations 
for 11 to 15 GHz observations of TMC-GB,
showing strong lines $HC_5N$ $J = 5 - 4$ (green), 
$HC_7N$ $J  =11 - 10$ (red), $HC_7N$ $J =12 - 11$ (blue) and 
$HC_9N$ $J =23 - 22$ (black).
These spectra are produced from the average of 103 minutes of 
daily test observations in 2006.
}
\end{figure}

\begin{figure}
\epsscale{.80}
\plotone{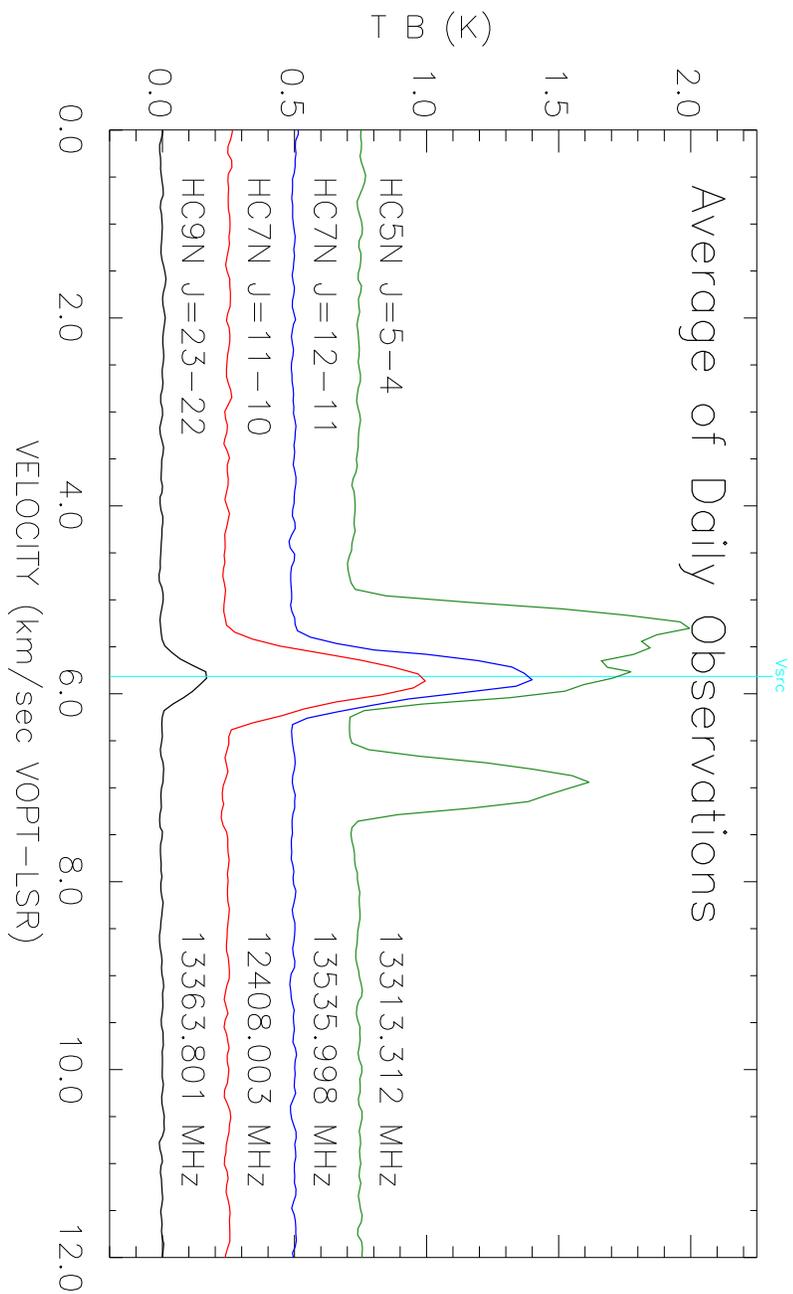}
\caption{Intensity versus velocity plot for $HC_9N$, $HC_7N$ and $HC_5N$
lines detected in the daily system check observations.  
The lines are offset by 0.25 K for clarity.  The plot shows the same data in the
same colors as shown in the previous plot.
}
\end{figure}

\begin{figure}
\epsscale{.80}
\plotone{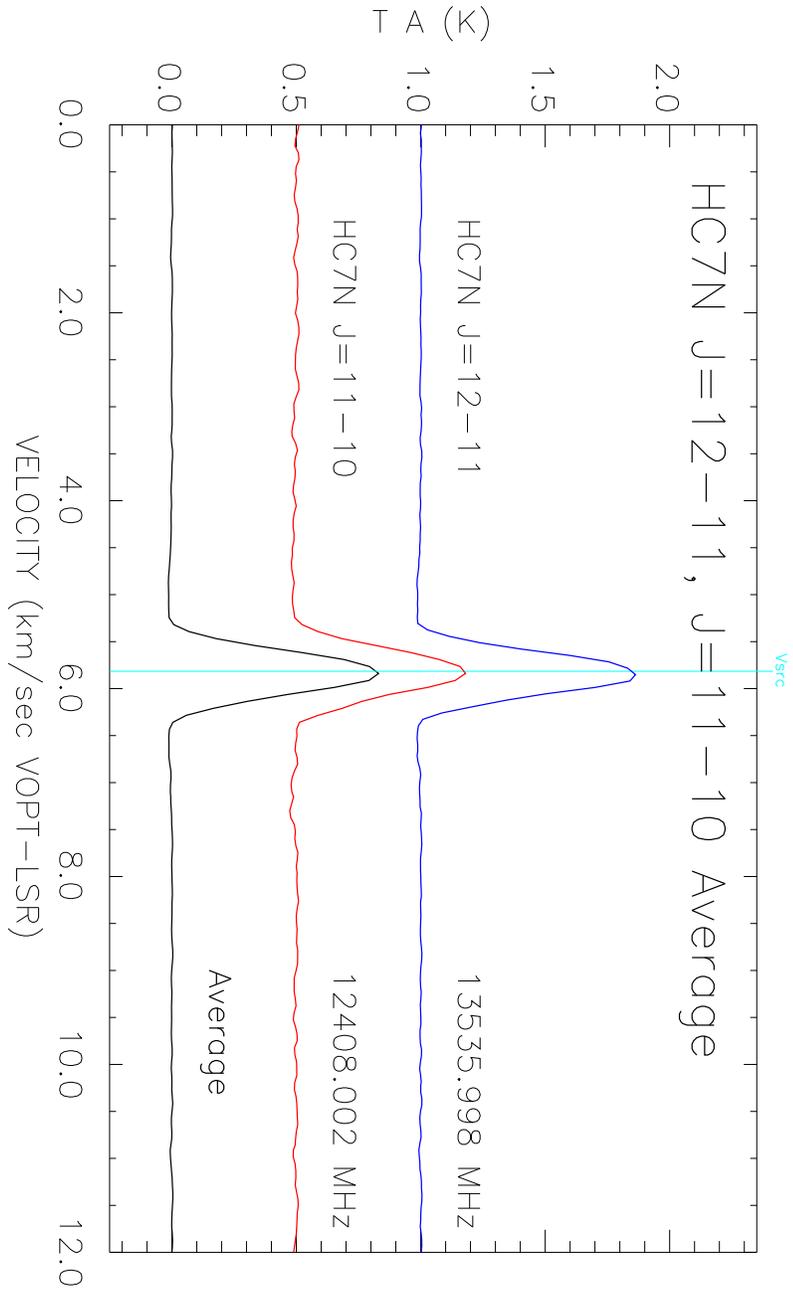}
\caption{Intensity versus velocity plot for the average of the 
$HC_7N$ $J = 12 - 11$ and $J = 11 - 10$  transitions at location \TMCG.
The weighted average of the two transitions is shown with zero offset.   
The two different $J' - J $ transitions are offset by 0.5 K. 
These spectra are produced from the average of 103 minutes of 
daily test observations in 2006.
}
\end{figure}

\begin{figure}
\epsscale{.80}
\plotone{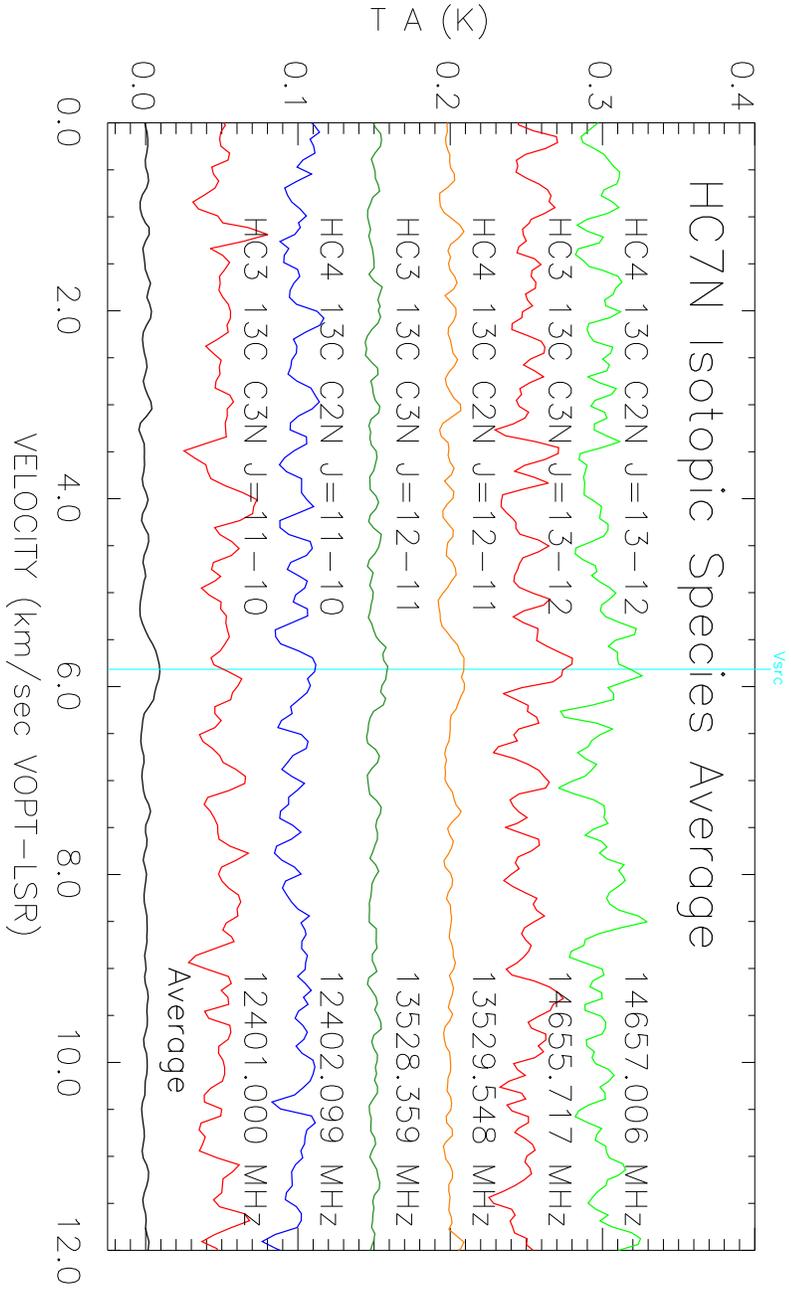}
\caption{Intensity versus velocity plot for a weighted average of $^{13}C$ isotopic 
species of $HC_7N$ listed in table 2.   
The weighted average is shown with
no intensity offset and successive lines are shown with 0.1 K offsets.
These spectra are produced from the average of 103 minutes of 
daily test observations in 2006.
}
\end{figure}

\begin{figure}
\epsscale{.80}
\plotone{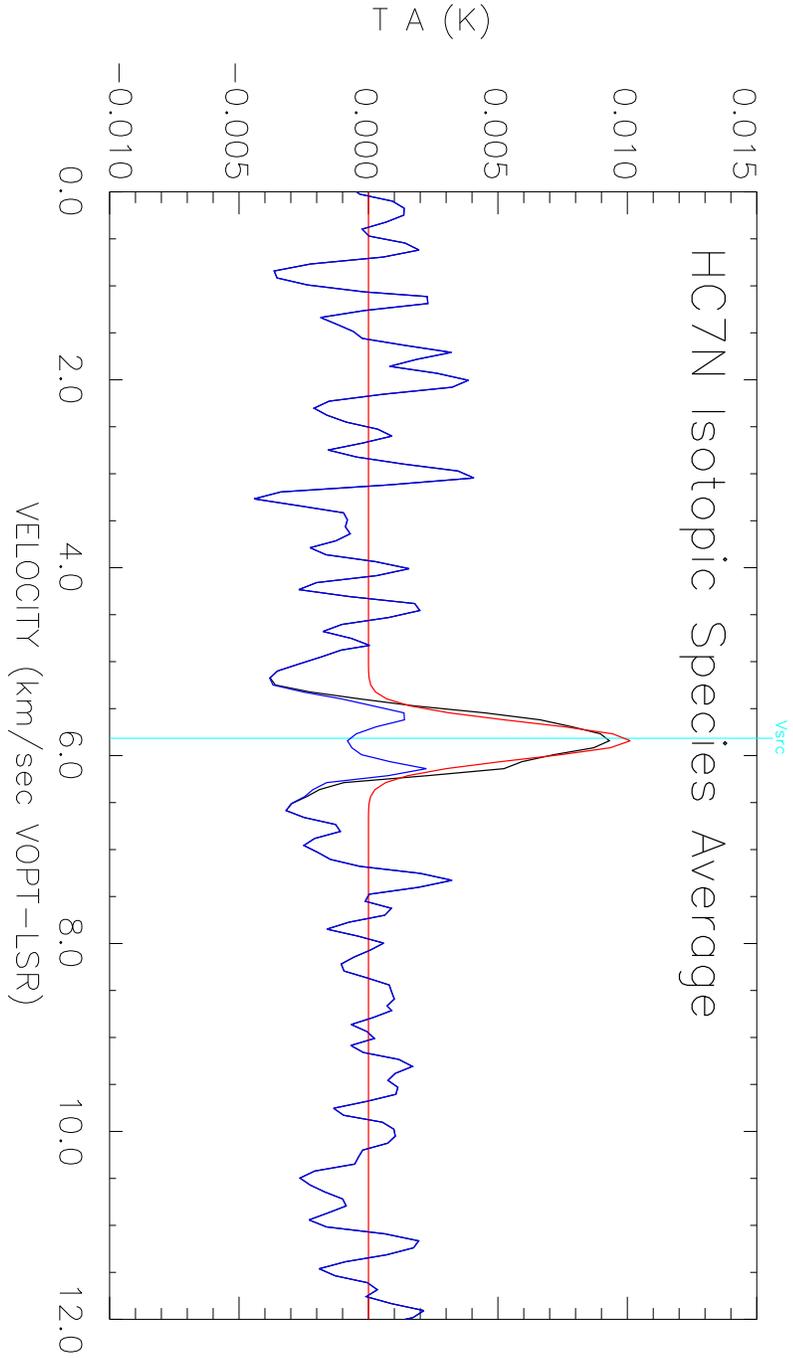}
\caption{Intensity versus velocity plot for the weighted average of isotopomers
of $HC_7N$.   Same weighted average data as the previous plot.
A single gaussian is fit to the average.   The difference between fit
and data is also shown.   The vertical line shows the reference velocity
5.815 km/sec LSR.
}
\end{figure}

\begin{figure}
\epsscale{.80}
\plotone{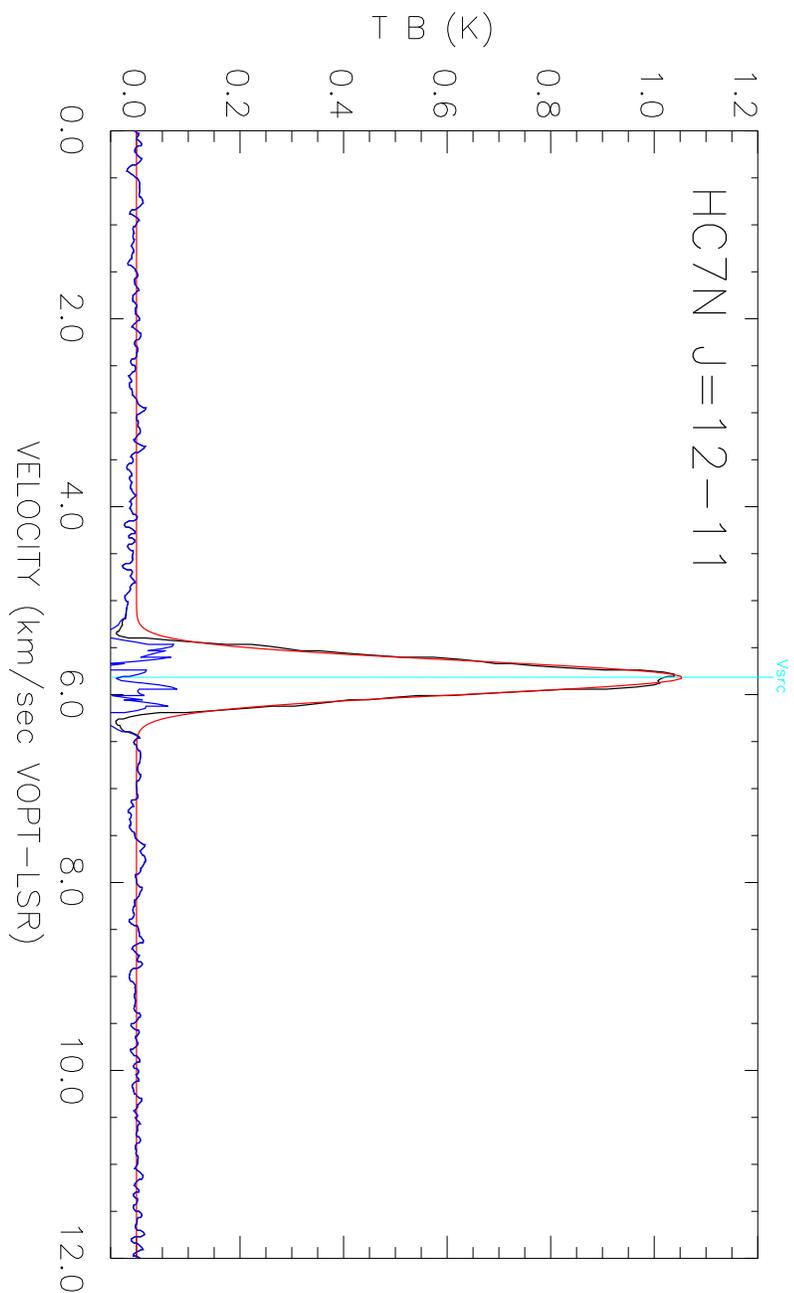}
\caption{Intensity versus velocity plot for 
$HC_7N$ $J = 12 - 11$  transition at location \TMCN.
The smooth red line shows a Gaussian fit to the data.  The
difference between fit and data is also shown.   
Spectrum is produced from an average of 87.5 minutes of observations in 
September 2006.
}
\end{figure}

\begin{figure}
\epsscale{.80}
\plotone{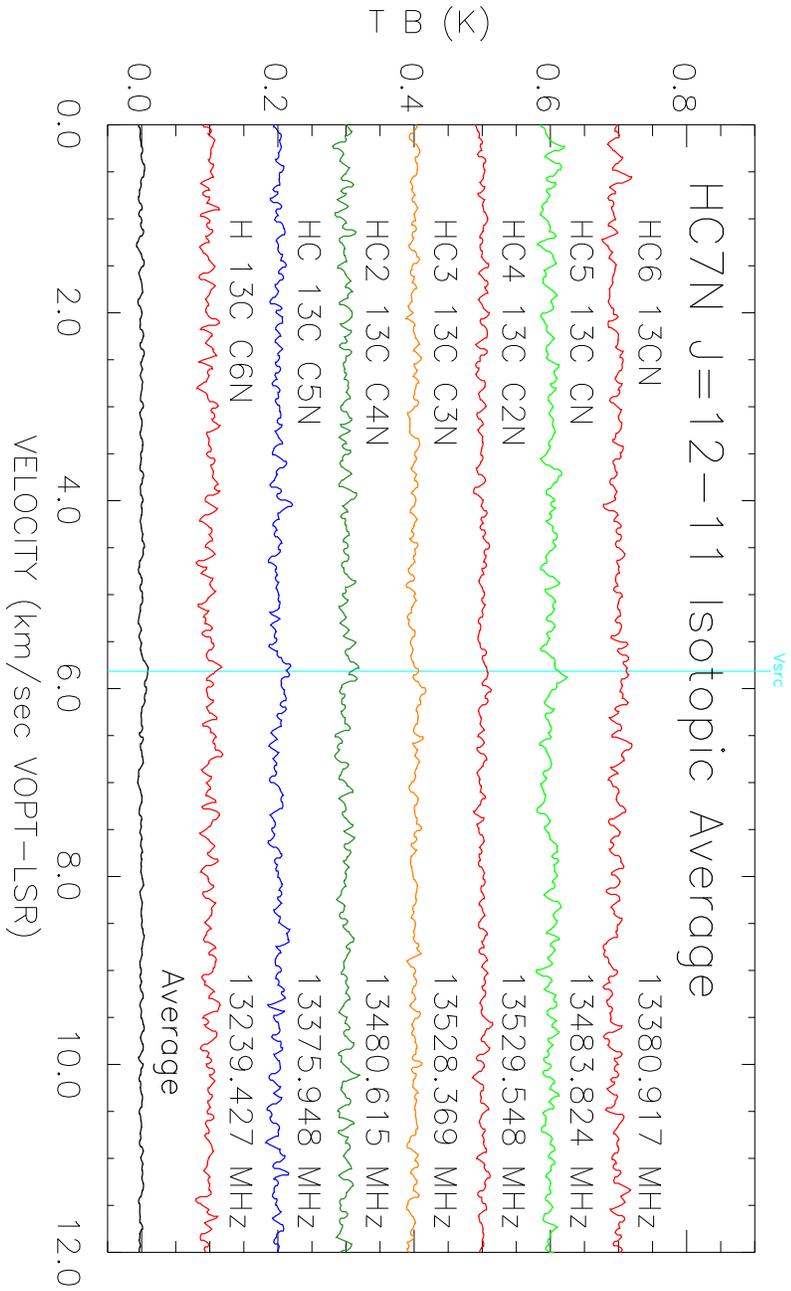}
\caption{Intensity versus velocity plot for each of the $^{13}C$ isotopomers
of $HC_7N$ at the location \TMCN.   The average of all
lines listed in table 3 is shown with zero offset, and 
the isotopomers are shown with 0.1 K offsets.   
The vertical line show the reference velocity.
Spectrum is produced from an average of observations in 
September 2006.
}
\end{figure}

\begin{figure}
\epsscale{.80}
\plotone{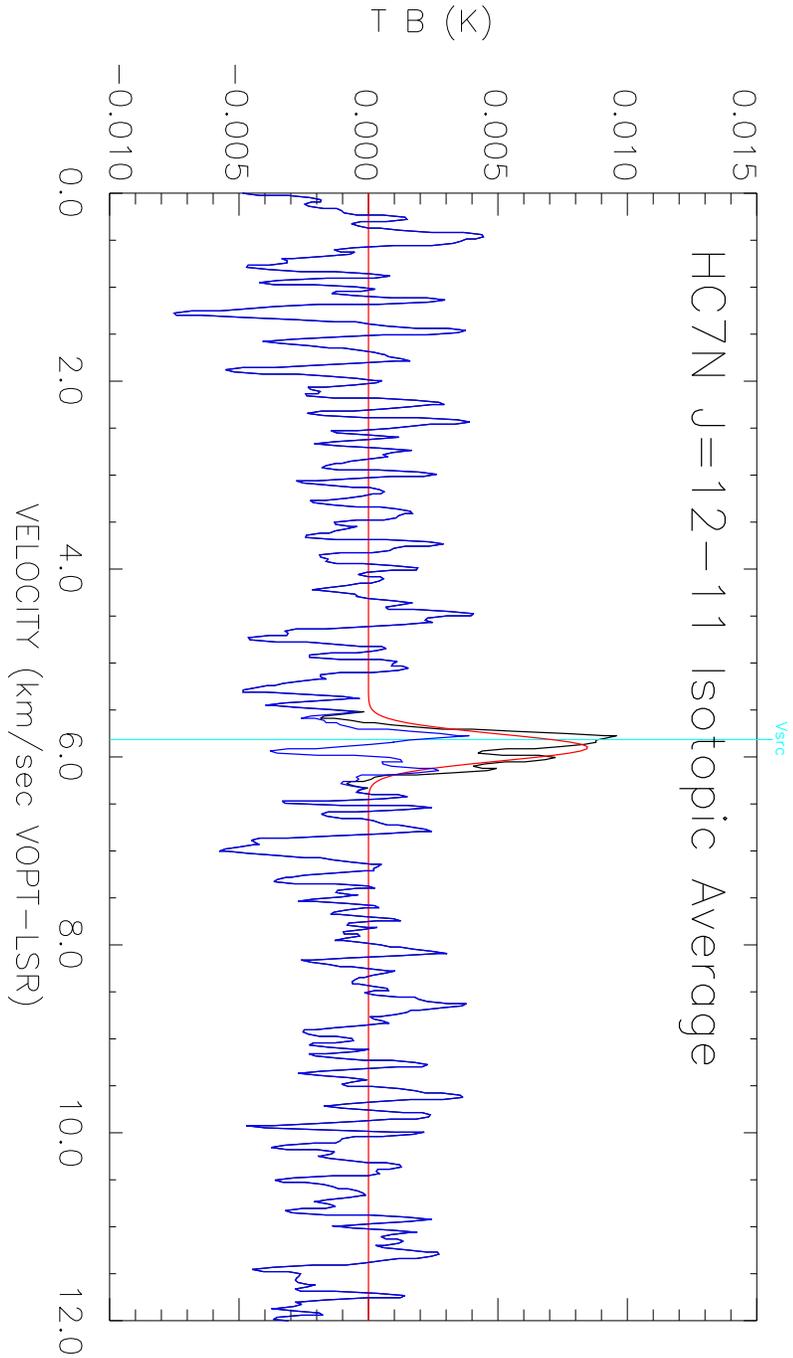}
\caption{Intensity versus velocity plot for the average of the $^{13}C$ isotopomers
of $HC_7N$ at location \TMCN.   
Same weighted average data as the previous plot.
The smooth red line shows a Gaussian fit to the data.
The difference between fit and data is also shown.
}
\end{figure}

\clearpage

\begin{center}
\begin{table}
\caption{Cyanopolyyne Lines Observed Daily}
\begin{tabular}{cccc}
\tableline\tableline
Molecule & Transition&                  &Rest Frequency \\
               & (J)            &      (F)       &(MHz)     \\
\tableline
$ HC_5N$ &  $5 -  4$&$4-3$&13313.259 \\
$ HC_5N$ &  $5 -  4$&$5-4$&13313.309 \\
$ HC_5N$ &  $5 -  4$&$6-5$&13313.331 \\
$ HC_7N$ & $11 - 10$&&12408.003 \\
$ HC_7N$ & $12 - 11$&&13535.998 \\
$ HC_9N$ & $23 - 22$&&13363.801 \\
\tableline
\end{tabular}
\end{table}
\end{center}
\begin{center}
\begin{table}
\begin{center}
\caption{Observed lines of rotational transitions of isotopic species of $HC_7N$
at location TMC-GB.
}
\begin{tabular}{lccccc}
\tableline\tableline
Molecule & Transition & Rest Frequency & $T_B$ & $v$  & $\Delta v$ FWHM  \\
               &  $J' - J$ & (MHz)     &  (K)   & (km/s)  & (km/s) \\
\tableline
$ HC_7N$                  & $11 - 10$&12408.003& 0.714\PM0.007 &  5.825\PM0.002 &  0.519\PM 0.003 \\
$ HC_7N$                  & $12 - 11$&13535.998& 0.970\PM0.005 &  5.819\PM0.002 &  0.459\PM  0.003 \\
$ HC_7N$                  &  average &  & 0.864\PM0.004 &  5.826\PM0.004 &   0.492\PM0.003 \\
\\
$ HC_3$$^{13}CC_3N$  & $11 - 10$&12401.000& $<$ 0.023 \\
$ HC_4$$^{13}CC_2N$  & $11 - 10$&12402.093& $<$ 0.023 \\
$ HC^{13}CC_5N$          & $12 - 11$&13375.947& $<$ 0.014 \\
$ HC_3$$^{13}CC_3N$  & $12 - 11$&13528.359& $<$ 0.017 \\
$ HC_4$$^{13}CC_2N$  & $12 - 11$&13529.548&  $<$ 0.017 \\
$ HC_6$$^{13}CN$        & $12 - 11$&13380.917&  $<$ 0.015 \\
\\
$^{13}C ~~ Isotopic~Ave$ & $12 - 11$   &   & 0.0099 \PM0.003& 5.782\PM0.054  & 0.52\PM0.13 \\
 \tableline
\end{tabular}
\end{center}
\end{table}
\end{center}

\begin{center}
\begin{table}
\begin{center}
\caption{Observed lines of rotational transitions of isotopic species of 
$HC_7N$ at location \TMCN.
}
\begin{tabular}{lccccc}
\tableline\tableline
Molecule & Transition & Rest Frequency & $T_B$ & $v$  & $\Delta v$ FWHM \\
               &  $J' - J$ & (MHz)     &  (K)   & (km/s)  & (km/s) \\
\tableline
$ HC_7N$             & $12 - 11$&13535.998& 1.101\PM 0.007 &   5.815\PM 0.001 &  0.426\PM     0.001 \\
$ DC_7N$             & $12 - 11$&13087.538& $< 0.019$ \\
$ HC_7$$^{15}N$      & $12 - 11$&13254.054& $< 0.019$ \\
\\
$ H^{13}CC_6N$       & $12 - 11$&13239.427& $< 0.019$ \\
$ HC^{13}CC_5N$      & $12 - 11$&13375.947& $< 0.019$ \\
$ HC_2$$^{13}CC_4N$  & $12 - 11$&13480.615& $< 0.019$ \\
$ HC_3$$^{13}CC_3N$  & $12 - 11$&13528.359& $< 0.015$ \\
$ HC_4$$^{13}CC_2N$  & $12 - 11$&13529.548&  $< 0.015$ \\
$ HC_5$$^{13}CCN$    & $12 - 11$&13484.825&  $< 0.019$ \\
$ HC_6$$^{13}CN$     & $12 - 11$&13380.917&  $< 0.019$ \\
\\
$^{13}C ~~ Isotopic~Ave$ & $12 - 11$& &  0.0084\PM0.0022 &   5.902 \PM 0.013  &  0.348 \PM 0.030  \\
\tableline
\end{tabular}
\end{center}
\end{table}
\end{center}

\end{document}